\documentclass[twoside, epsfig]{article}

\usepackage[round]{natbib}

\usepackage{graphicx}
\usepackage{color}
\usepackage{array}
\usepackage{lscape}
\usepackage{rotating}
\usepackage{longtable}
\usepackage{multirow}
\usepackage{fancyhdr}
\usepackage[colorlinks=true,urlcolor=blue,citecolor=black,bookmarks=false]{hyperref}
\usepackage{gensymb}

\font\fhead=cmr8
\font\ftitle=cmssbx14
\font\fabs=cmr10

\usepackage{amssymb}
\usepackage{makecell}
\usepackage{enumitem}

\def\MyHead#1 {
\pagestyle{fancy}
\renewcommand{\headrulewidth}{0pt}
\lhead{{\fhead #1}}
}

\def\MyTitle#1 {
  {\centerline {\ftitle #1} \vskip 0.5cm plus 0.4cm minus 0.2cm}}

\def\MyTitleTwo[#1]#2 {
  {\centerline {\ftitle #1} \vskip 0.3cm
   \centerline {\ftitle #2} \vskip 0.5cm plus 0.4cm minus 0.2cm}
}

\def\MyAuth#1 {
 \centerline {#1}
 \vskip 0.5cm
}

\newcounter{authno}
\def\MyInst#1{ {\centering \small \addtocounter{authno}{1} ${\arabic{authno}})$ #1
        \vskip 1mm }\large}

\def\MyAbstract#1{
    \vskip 0.5cm
    \begin{center}
    \begin{tabular}{p{0.9\textwidth}}
    {\fabs {\bf Abstract:} #1 }\\
    \end{tabular}
    \end{center}
}

\def\MyAIusage#1{
    \vskip 0.5cm
    \noindent{\fabs{\bf Disclosure of AI usage:} #1 }\\
}

\def\MyAcknowledgements#1{
    \vskip 0.5cm
    \noindent{\fabs{\bf Acknowledgements:} #1 }\\
}

\def\MyBibcode#1{
    \href{https://ui.adsabs.harvard.edu/abs/#1}{\underline{#1}}
}

\def\MyLink#1{
     \href{#1}{\underline{#1}}
}

\begin{document}

\MyHead{September 2026}

\MyTitleTwo[V405 Draconis: A Deep Contact Binary with a Tertiary Companion]{and a Quadruple System Candidate}

\MyAuth{Maksym Yu. Pyatnytskyy$^1$ and Ivan L. Andronov$^2$}
\MyInst{Private Observatory ”Osokorky” Kyiv, Ukraine, {\tt \href{mailto:pmak@osokorky-observatory.com}{pmak@osokorky-observatory.com}}}

\MyInst{Department "Higher Mathematics", Odesa National Maritime University, Odesa, Ukraine}

\MyAbstract{
We continued our analysis of the O–C diagram of the EW-type eclipsing binary V405 Dra using recently available data. We refined the orbital parameters of the third body responsible for the short-period oscillations in the O–C curve, obtaining $P_3=513.12(20)\,\mathrm{d}$, $a_{12}\sin i_3=0.3862(15)\,\mathrm{AU}$, and $e_3=0.1690(48)$. Assuming a total mass of $1.56\,M_\odot$ for the eclipsing binary, the minimum mass of the third body is estimated to be about $0.5\,M_\odot$. We also investigated the complex long-term O–C trend and found that it may be partially explained by the light-time effect caused by a possible fourth body, with a characteristic period of about 23.5 yr and a minimum mass of about $0.9\,M_\odot$. However, the fourth-body model cannot fully reproduce the observed long-term variation, particularly the earliest data, and the nature of part of the long-term trend remains unexplained.
}

\section{Introduction}

V405 Draconis is an eclipsing binary of the EW (W UMa) type. We have been observing V405 Dra since 2021. The system has shown complex changes in its O–C diagram \citep{mpyat2022, mpyat2024}. We attributed the periodic variations in the O–C curve to the presence of a faint third component \citep{mpyat2022}. However, the nature of the long-term trend, which cannot be described by a simple parabolic approximation, remains unclear.

\citet{li2024} interpreted part of the long-term trend as a result of unusually rapid mass transfer between the components; however, their interpretation does not account for other parts of the O–C curve. Recent years have provided additional data that allow us to improve the estimates of the third body's parameters but also reveal an even more complex long-term trend.

In the present work, we report the results of our analysis of the third body and discuss possible interpretations of the long-term trend. We use our own observations, observations from the International AAVSO Database (AID) \citep{kloppenborg2023}, and TESS data.

\section{Observation and data reduction}

We conducted observations in the Osokorky neighborhood, Kyiv, Ukraine. We used a 150 mm Newtonian telescope with a focal length of 750 mm, equipped with a ZWO ASI183MM Pro CMOS camera and a Johnson V photometric filter. The observations cover the period from August 2021 to May 2024.

We used the standard image calibration procedure, using dark, flat, and dark-flat frames, which were collected each observing night immediately after gathering the light frames. Calibration was performed with the MaxFITStoolkit package\footnote{\MyLink{https://github.com/mpyat2/MaxFITStoolkit}}. Photometry was conducted using the AstroImageJ package \citep{collins2017}. All data were uploaded to the AID.

\section{Data analysis}

To build the O–C diagram covering the longest possible period of time, we used different sources of information, including our observations, data from the NSVS \citep{wozniak2004}, SuperWASP \citep{butters2010}, ASAS-SN \citep{kochanek2017}, and TESS \citep{ricker2014} surveys, published times of minima \citep{diethelm2006, diethelm2007}, and observations found in the AID.

To derive times of minima (TOMs) from the observational data, we used the Wall Supported Line (WSL) algorithm \citep{andrych2017, andrych2020}. This algorithm works well for minima with nearly flat bottoms, which are characteristic of totally eclipsing binary stars like our target. For the NSVS, SuperWASP, and ASAS-SN data, we first created a folded light curve for each observational year because the data from these surveys are sparse, and then determined the positions of the minima from the folded light curves.

\begin{figure}[htbp]
\centering
\includegraphics[width=15cm]{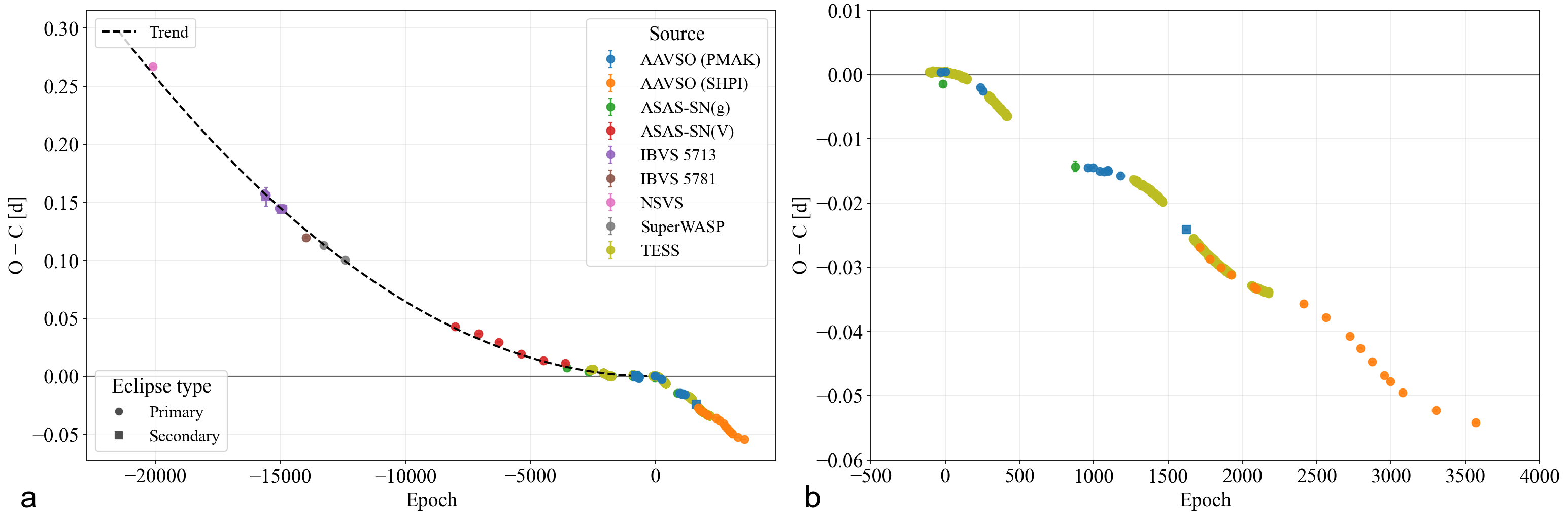}
\caption{The O–C diagram calculated with a period of 0.4130518 days and initial epoch $BJD_{TDB} = 2459763.3895$.}
\label{fig_OC_vs_epoch}
\end{figure}

Figure~\ref{fig_OC_vs_epoch} shows the O–C diagram including recent data. We used a period of 0.4130518 d and an initial epoch of $BJD_{TDB} = 2459763.3895$ from \citep{mpyat2024}. A parabolic trend reasonably describes the O–C points before epoch 0 (i.e., up to $BJD_{TDB} = 2459763.3895$). The apparent period increase rate is $dP/dt = 1.14(1)\times10^{-6} d^{-1}$ for this interval. Then, the trend abruptly declines.

In the part of the O–C diagram where the points are fairly densely spaced, periodic oscillations are clearly visible. We previously suggested that these periodic changes could be attributed to the light-travel-time effect caused by the presence of a third body \citep{mpyat2022}.

\subsection{Analysis of the O–C Periodic Oscillation}

Analysis of the periodic O–C oscillations, which are most likely caused by the presence of a third body, allows us to determine the orbital parameters of the third body. Under certain assumptions, its mass can also be estimated.

For the analysis, we used TESS and AAVSO data only (including our observations), since they contain well-sampled light curves in the vicinity of minima.

The oscillations are accompanied by a trend of complex shape. To account for this trend, we used a combined model implemented in a custom Python script. The model consists of a polynomial of selectable order, which phenomenologically describes the long-term trend, and a component describing the oscillations. To model the oscillations, we used the {\it ocfit.LiTE} function from the {\it OCFit} package \citep{gaidos2019, gaidos2023}. This function calculates the O–C variation caused by the light-time effect for a Keplerian orbit, using the projected semi-major axis $a_{12} \sin i_3$, eccentricity $e$, argument of periastron $\omega_3$, time of periastron passage $T_{0_3}$, and orbital period $P_3$ as model parameters. Using the {\it scipy.optimize} module, the script searched for the optimal values of these parameters. The orbital parameters were optimized by minimizing the chi-square statistic, which measures the weighted differences between the observed O–C values and the combined model. For each trial set of orbital parameters, the polynomial coefficients were determined by weighted linear least squares.

We tested the model using different degrees of the polynomial trend, up to degree 15. We found that the ninth-degree polynomial is the lowest-degree polynomial for which the model adequately describes the long-term trend without absorbing part of the oscillating component. The final solution was refined and the parameter uncertainties were estimated using the Markov Chain Monte Carlo method implemented in the {\it emcee} package \citep{foreman-mackey2013}.

\begin{figure}[htbp]
\centering
\includegraphics[width=15cm]{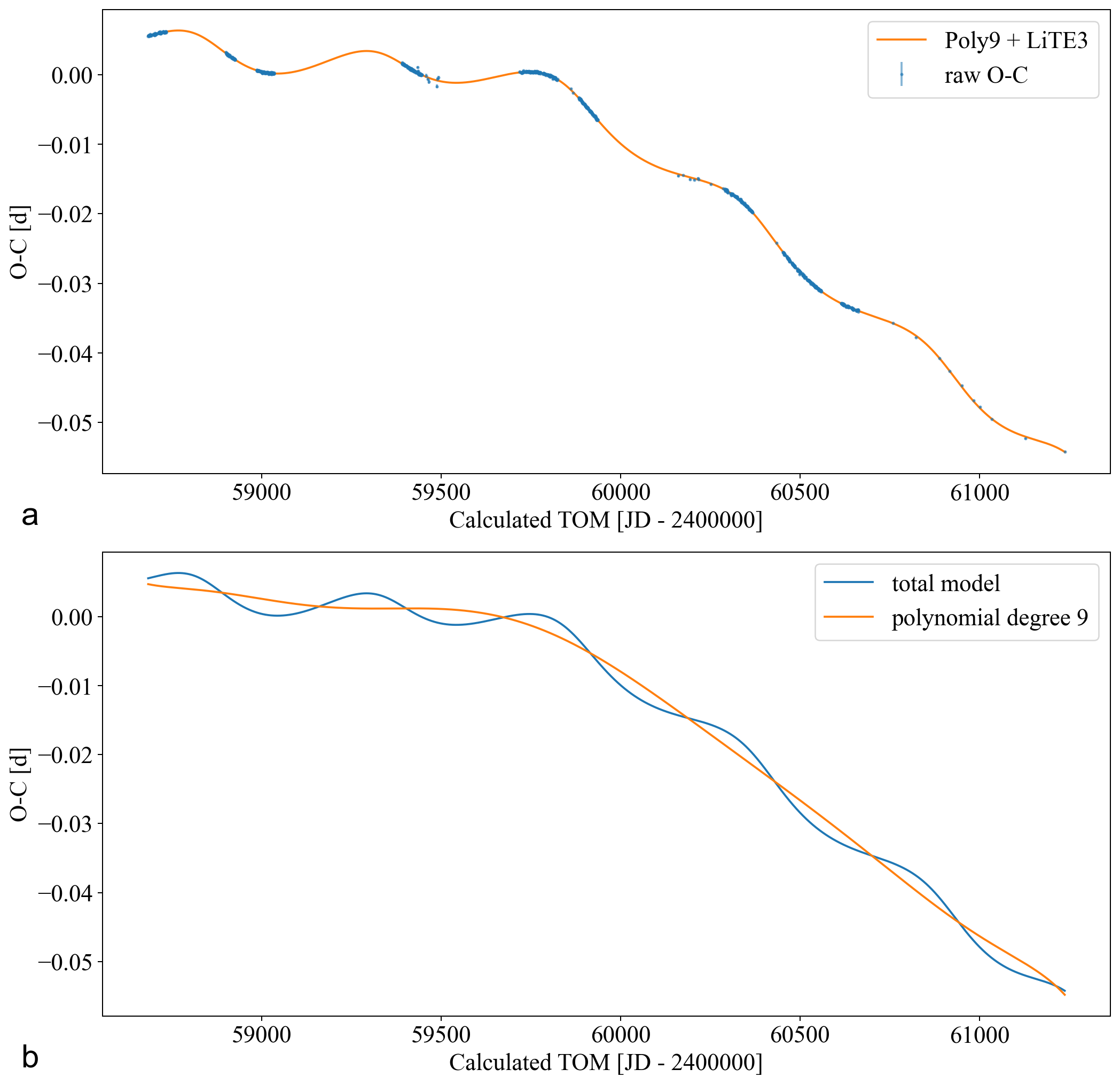}
\caption{(a) The best-fitting model; (b) the full model and its polynomial component.}
\label{fig_OC_trend}
\end{figure}

Figure~\ref{fig_OC_trend} shows the best approximation. The upper panel shows the approximation along with the O–C points; the lower panel depicts the full model and its polynomial part separately.

\begin{figure}[htbp]
\centering
\includegraphics[width=15cm]{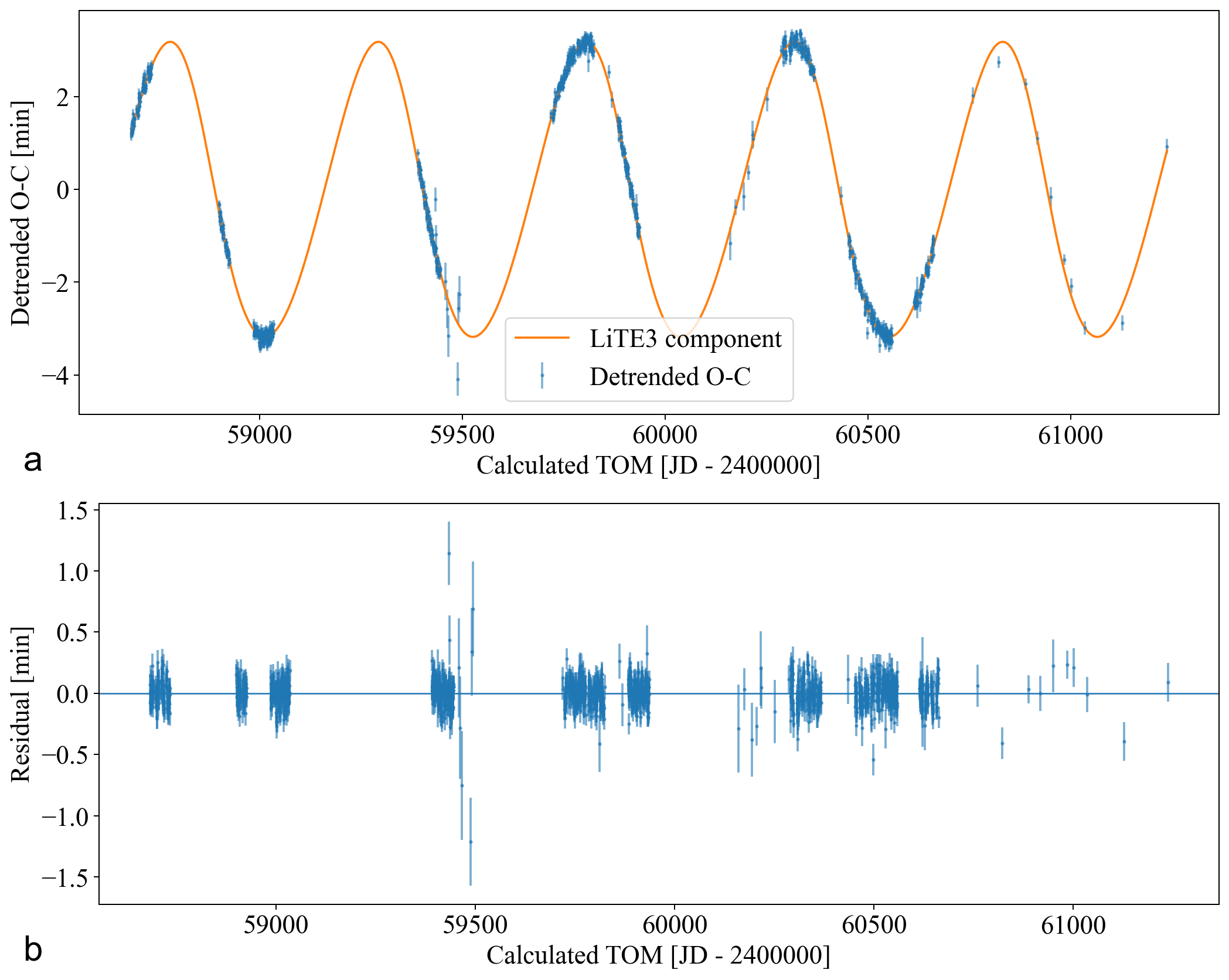}
\caption{(a) The component of the model associated with the light-time effect, along with the O–C points; (b) residuals.}
\label{fig_OC_light_component}
\end{figure}

Figure~\ref{fig_OC_light_component} shows the detrended O–C curve and the light-time effect component of the model associated with the third body. The lower panel shows the residuals.

The obtained orbital parameters are listed in Table~\ref{tab1}.

We also estimated the mass of the third body by adopting an approximate mass for the EW system derived from the empirical relation between orbital period and total mass. V405 Dra is likely to be a deep, low-mass-ratio overcontact binary: \citet{li2024} obtained a mass ratio of $q=0.175$ and a fill-out factor of about 69\%. We therefore used the period--total-mass relation derived by \citet{yang2015} for deep, low-mass-ratio overcontact binaries. This relation yields a total mass of approximately $1.56\,M_\odot$ for the EW system, which we adopted in estimating the mass of the third component. The estimated masses corresponding to three different orbital inclinations are listed in Table~\ref{tab1}.

It is worth noting that the obtained orbital eccentricity is considerably higher than that reported by \citet{li2024} (0.085). In addition, the projected semi-major axis of the eclipsing binary, 0.3862 AU, is lower than the value of 0.56 AU reported by \citet{li2024}, resulting in smaller estimates of the third body's mass.


\begin{table}
\caption{Orbital parameters of the third body and its estimated masses for different inclinations}\vspace{3mm}  
\centering
\begin{tabular}{lcl}
\hline
$P_3$(d) & 513.12(20) \\
$a_{12} \sin i_3$(AU) & 0.3862(15) \\
$e_3$ & 0.1690(48) \\
$\omega_3$(\degree) & 148.5(1.7) \\
$T_{0_3}$($BJD_{TDB}$) & 2459364.3(2.2)  \\
$f(m)(M_{\odot})$ & 0.02918(33)\\
\hline
Assumed $M_{12}(M_{\odot})$ & 1.56 \\
$M_3(i_3 = 90\degree)$ ($M_{\odot}$) & 0.50 \\
$M_3(i_3 = 75\degree)$ ($M_{\odot}$) & 0.52 \\
$M_3(i_3 = 60\degree)$ ($M_{\odot}$) & 0.59 \\
\hline
\end{tabular}\label{tab1}
\end{table}

\subsection{Long-term trend}

\begin{figure}[htbp]
\centering
\includegraphics[width=15cm]{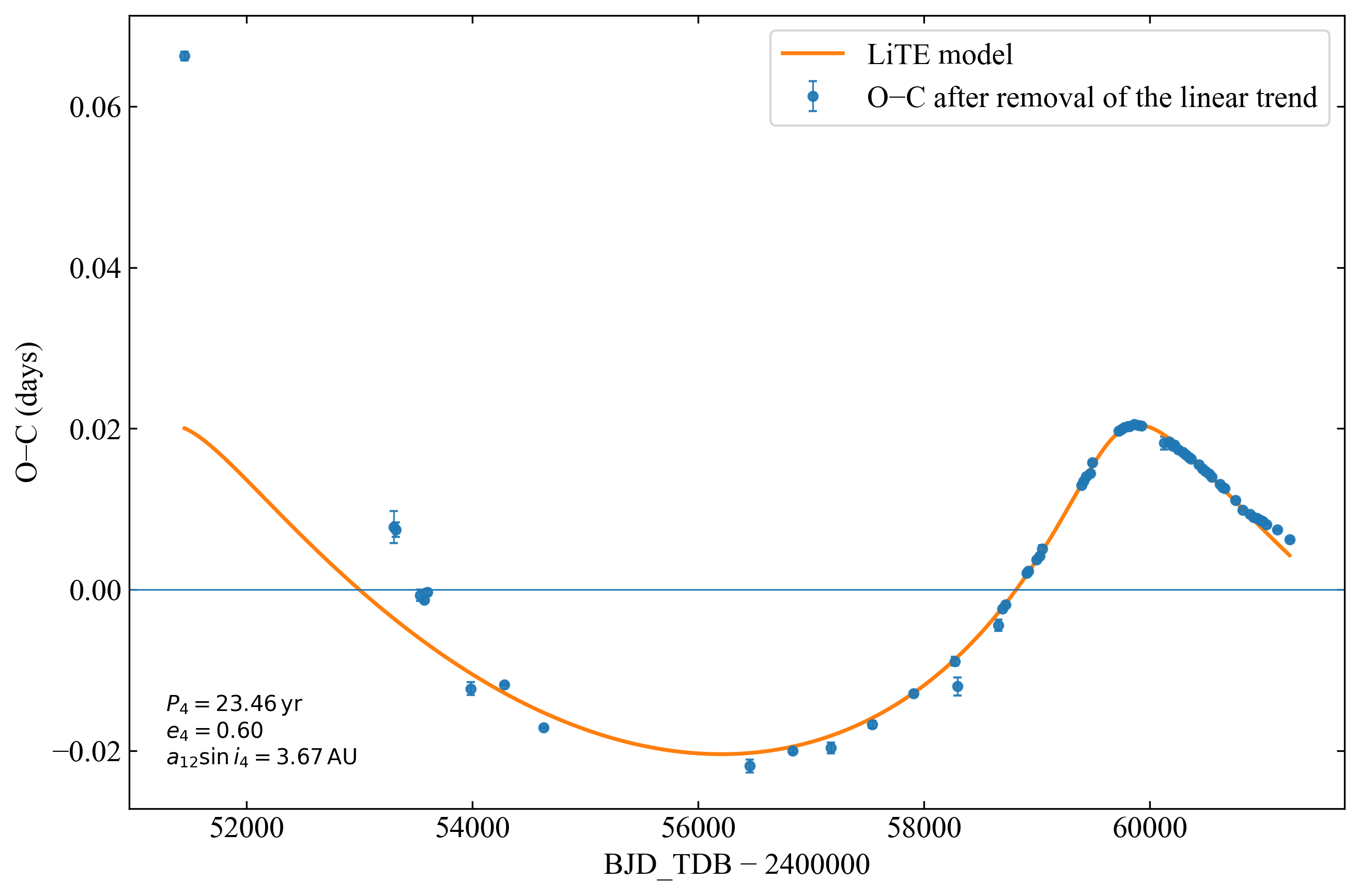}
\caption{The best-fit approximation of the long-term O–C trend, assuming a light-time effect caused by a possible fourth body.}
\label{fig_V405_Dra_LiTE4_detrended}
\end{figure}

We removed the short-period oscillations attributed to the third body from the O–C data. We then applied another light-time-effect model, combined with a linear trend, to the residual O–C curve. Before fitting, the data were binned into 30-day intervals to reduce the predominance of the densely sampled TESS data. The best-fit solution yielded a characteristic period of $P_4 \approx 23.5$ yr, an eccentricity of $e_4 \approx 0.6$, and a projected semi-major axis of $a_{12}\sin i_4 \approx 3.7$ AU. The corresponding approximation, after removal of the fitted linear trend, is shown in Fig.~\ref{fig_V405_Dra_LiTE4_detrended}. It is seen that this solution does not adequately describe the entire O–C curve. Although it reproduces the data at $BJD_{\rm TDB} > 2454000$ reasonably well, the earlier points show considerable deviations from the model. Thus, the presence of a fourth body may explain part of the observed long-term variation, but the available data do not allow us to establish this unequivocally.

Based on the preliminary orbital parameters given above, and adopting $M_{12}=1.56\,M_\odot$ for the eclipsing binary and a minimum mass of $M_3=0.50\,M_\odot$ for the third body, the minimum mass of the possible fourth body is estimated to be approximately $0.9\,M_\odot$.

The presence of a possible fourth body with a mass comparable to that of the Sun may explain a considerable fraction of the additional light found by \citet{li2024}. The third body, with an estimated mass of about $0.5-0.6\,M_\odot$, is expected to make only a small contribution to the total luminosity of the system. However, the possible fourth body has a minimum estimated mass of about $0.9\,M_\odot$ and, if it is a main-sequence star, may account for a considerable part of the remaining additional light. For lower orbital inclinations, its estimated mass, and consequently its expected luminosity, would be higher.

\section{Conclusions}

We analyzed the O–C curve of V405 Draconis over a time span of more than 25 years. We found that, in addition to a complex long-term trend that cannot be explained by a monotonic period change, well-defined short-period oscillations are present in the O–C curve. These oscillations, with a period of about 1.4 yr, can be well explained by the light-time effect caused by a third body with a minimum mass of about $0.5\,M_\odot$. The long-term variation may be partially explained by the presence of a fourth body with a minimum mass of about $0.9\,M_\odot$. Such a body, if it is a main-sequence star, may also account for a considerable fraction of the additional light previously reported for this system \citep{li2024}. However, the fourth-body model cannot fully reproduce the observed long-term O–C variation, particularly the earliest data. Therefore, the presence of a fourth body provides a possible explanation for a considerable part of the long-term variation, although some features of the O–C curve remain unexplained. Additional long-term monitoring of V405 Dra is required to obtain more information that may help to explain its complex O–C behavior.

\setcounter{secnumdepth}{0}

\MyAIusage{
During the preparation of this manuscript, the authors occasionally utilized artificial intelligence tools to assist with text editing, language refinement, and the generation of Python scripts for data analysis and figure preparation. All AI-generated content and code were thoroughly reviewed, verified, and finalized by the authors, who take full responsibility for the contents of the publication.
}
\\

\MyAcknowledgements{
This paper includes data collected by the Transiting Exoplanet Survey Satellite (TESS) mission, which are publicly available from the Mikulski Archive for Space Telescopes (MAST). Funding for the TESS mission is provided by NASA's Science Mission Directorate. This research has made use of data products from the TESS Science Processing Operations Center (SPOC).
This publication makes use of data from the Northern Sky Variability Survey (NSVS), created jointly by the Los Alamos National Laboratory and the University of Michigan. The NSVS was funded by the U.S. Department of Energy, the National Aeronautics and Space Administration, and the National Science Foundation.
This paper makes use of data from DR1 of the WASP data as provided by the WASP consortium, and computational resources supplied by the project “e-Infrastruktura CZ” (e-INFRA CZ LM2018140), supported by the Ministry of Education, Youth and Sports of the Czech Republic.
This research has also made use of data from the All-Sky Automated Survey for Supernovae (ASAS-SN). We thank Las Cumbres Observatory and its staff for their continued support of ASAS-SN.
We gratefully acknowledge the contributions of the AAVSO observer community, whose photometric data and metadata resources were used in this study and made available through the AAVSO's scientific archives. In particular, we acknowledge the observations of V405 Draconis contributed to the AAVSO International Database by observer SHPI.
} 
\\

\end{document}